\documentclass[a4paper,aip,apl,amsmath,amssymb,reprint,superscriptaddress]{revtex4-1}
\pdfoutput=1
\usepackage{graphicx}
\usepackage{amsmath}
\usepackage{dcolumn}
\usepackage{bm}
\usepackage{bbold}
\usepackage{color}
\usepackage{amsfonts}
\usepackage{amssymb}
\usepackage{mathrsfs}
\usepackage{tabularx}
\usepackage{braket}
\usepackage{mathtools}
\usepackage{soul}			% For highlighting text, use command \hl
\usepackage{multirow}
\usepackage{epstopdf}

\usepackage{caption}
\usepackage{subcaption}
\usepackage[font=small]{caption}

\begin{document} 
\title{Determination of Niobium Cavity Magnetic Field Screening via a Dispersively Hybridized Magnonic Sensor}

\author{Graeme Flower}
\thanks{Author to whom correspondence should be addressed: 21302579@student.uwa.edu.au}
\affiliation{ARC Centre of Excellence for Engineered Quantum Systems, Department of Physics, University of Western Australia, 35 Stirling Highway, Crawley WA 6009, Australia}

\author{Benjamin McAllister}
\affiliation{ARC Centre of Excellence for Engineered Quantum Systems, Department of Physics, University of Western Australia, 35 Stirling Highway, Crawley WA 6009, Australia}

\author{Maxim Goryachev}
\affiliation{ARC Centre of Excellence for Engineered Quantum Systems, Department of Physics, University of Western Australia, 35 Stirling Highway, Crawley WA 6009, Australia}

\author{Michael E. Tobar}
\email{michael.tobar@uwa.edu.au}
\affiliation{ARC Centre of Excellence for Engineered Quantum Systems, Department of Physics, University of Western Australia, 35 Stirling Highway, Crawley WA 6009, Australia}

\date{\today}

%%%%%%%%%%%%%%%%%%%%%%%%%%%%%%%%%%%%%%%%%%%%%%%%%%%%

\begin{abstract}
A method for determining the internal DC magnetic field inside a superconducting cavity is presented. The method relies on the relationship between magnetic field and frequency of the Kittel mode of a ferrimagnetic sphere, hybridised in the dispersive regime of the superconducting cavity. Results were used to experimentally determine the level of screening a superconducting Nb cavity provides as it changes from perfect diamagnetism to no screening. Two cavity geometries were tested, a cylinder and single post re-entrant cavity. Both demonstrated a consistent value of field that enters the cavity, expected to be the superheating critical field. Hysteresis in the screened field during ramp up and ramp down of the external magnetic field due to trapped vortices was also observed. Some abnormal behaviour was observed in the cylindrical cavity in the form of plateaus in the internal field above the first critical field, and we discuss the potential origin of this behaviour. The measurement approach would be a useful diagnostic for axion dark matter searches, which plan on using superconducting materials but need to know precisely the internal magnetic field.

\end{abstract}
\maketitle

%\section*{Introduction}
Superconducting cavities are of great benefit for many scientific and engineering disciplines, which require low loss and high quality-factor resonators. For example, frequency metrology \cite{superclock}, particle accelerators \cite{SRFfermi,neutronsource,linearAccelerators,LHC} and tests of fundamental physics\cite{axionrfsuper}.  Axion haloscope dark matter experiments, which rely on low loss microwave cavities in high DC magnetic fields \cite{fom}, may be enhanced by the use of type II superconductors operated between the first and second critical fields, thus allowing flux into the cavity but retaining some superconductivity - interest in this has recently grown \cite{kaistYBCO}. Such experiments require that the DC magnetic field inside the cavity be both large and precisely known for a sensitive experiment. Thus, a system to measure the internal field is necessary for the calibration and development of such experiments, and it must be verified that large DC magnetic fields can penetrate superconducting cavities whilst retaining high quality factors. Any sensor to measure the internal field would need to operate in vacuum, and at cryogenic temperatures. Additionally, operating between the first and second critical field would open up the use of type II superconductors for improving loss in hybrid quantum systems based on cavities and magnetic materials \cite{CMP_Life, QuantaMagnons, QubitMagnon, magnonMeetsQubit, microlightconv1, Magnonqed1, Magnonqed2, Magnonqed3} including ferrimagnetic-axion haloscopes \cite{broadeningAxionCMP, magnonhalo}. This work presents a method for making measurements of the internal DC magnetic field inside superconducting microwave cavities immersed in an external magnetic field. 

Superconductors interact with nearby magnetic fields. Type I and II superconductors below their critical field (or first critical field for type II), behave as perfect diamagnets, generating supercurrents by the Meissner effect \cite{meissner} to perfectly screen external magnetic fields such that none is contained internally. Type I superconductors, fully explained by Bardeen-Cooper-Schrieffer (BCS) theory\cite{BCS}, will undergo a phase transition returning to normal conductivity above their critical field, allowing external DC fields to pass through with negligible screening. Type II superconductors on the other hand, above their first critical field will enter a mixed state, allowing some field to pass through. This occurs via formation of superconducting vortices surrounding a small region of normal conductor, with each vortex allowing a flux quantum to penetrate. As the field approaches the second critical field, all external flux can penetrate the superconductor as it forms an Abrikosov flux lattice, determined by Ginzburg-Landau (GL) theory \cite{abrikosov}. Above this second critical field, the system returns to its normal state.

Whether a superconductor is type I or II is determined by the Ginzburg-Landau parameter, $\kappa$, where for type I $\kappa<1/\sqrt{2}$, and type II $\kappa>1/\sqrt{2}$. A further distinction can be made in type II superconductors, where if $\kappa\sim1/\sqrt{2}$, it is denoted Type II/1. Such materials, including very pure Niobium, can form an intermediate mixed state (IMS) above the first critical field \cite{vortexMatterTheory1, typeII1,IMSNb}, in which the superconductor forms some domains in the Meissner state, and other domains containing vortices. Vortex matter can appear in type II superconductors in the mixed phase, where vortices become pinned in place forming solid like states. Vortex motion can produce an additionally channel for losses. This includes vibration in all cases and flow in vortex liquids. Flux pinning, where vortices can become trapped by defects or impurities in the crystal, can provide a barrier for vortex motion; however, even in vortex glass states, flux creep allows loss to occur\cite{annett}. Vortex matter, typically arising due to thermal fluctuations compared to the mean field GL theory, are primarily found in high temperature superconductors, however, they are also expected\cite{vortexMatterTheory1} and observed\cite{vortexMatter1,vortexMatter2} in low temperature superconductors like Niobium. Heat treatment of the material also significantly affects how materials like Niobium will trap or expel magnetic flux as they cool\cite{ExpelFlux1, ExpelFlux2, superheatingNb2}.

%\section{LiFe Magnons as Magnetic field sensor}
\begin{figure}[h!]
	\includegraphics[width=0.6\columnwidth]{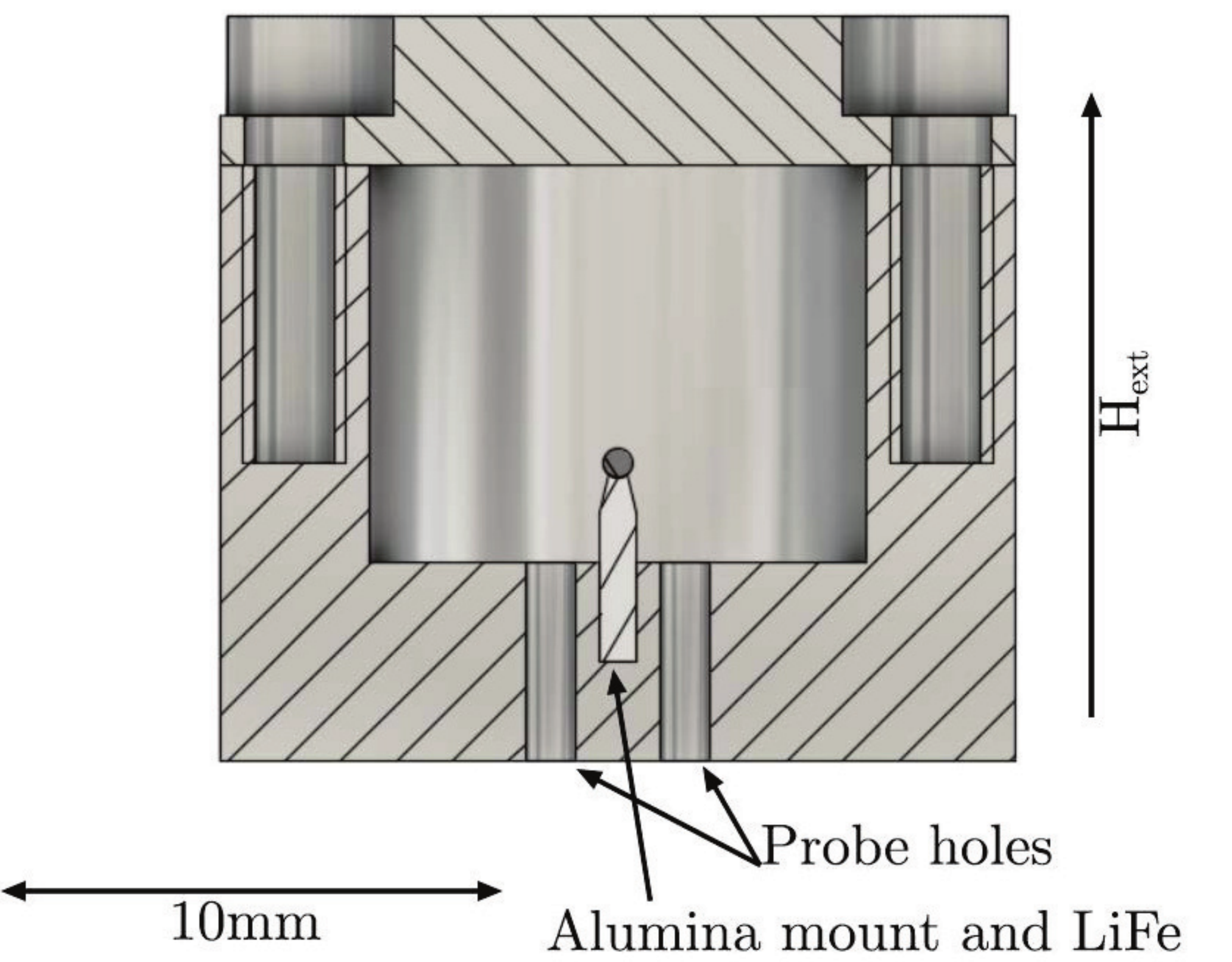}
	\caption{Cross-section of the Cylindrical cavity with a spherical LiFe magnetic field sensor inserted.}
	\label{cylndraw}
\end{figure}

Ferrimagnetic materials contain a collection of spins, in which the collective excitations are called magnons. Spherical geometries of ferrimagnetic materials such as lithium ferrite (LiFe), have a uniform precession mode known as a Kittel mode, which has a linear relation between applied magnetic field, $B_{DC}$, and its mode frequency, $\omega_{m}$ \cite{cavitymagnon},
\begin{equation}\label{FtoB}
\omega_m = \frac{g_\text{eff}\mu_B}{\hbar}(B_{DC}+B_\text{off}),
\end{equation}
where $\mu_B$ is the Bohr magneton, $B_\text{off}$ is an offset field typically due to magneto-crystalline anisotropies and $g_\text{eff}$ is the effective Land\'e g-factor. Thus the procedure for its use as a magnetic field sensor was simple. The sensor was placed inside a normal conducting cavity, as shown in fig.\ref{cylndraw}, and the Kittel mode frequency was measured over the desired magnetic field range. A fit for calibration was implemented, which related DC magnetic field incident on the LiFe, to the magnon Kittel mode frequency. The field incident on the sphere was taken to be the same as the external applied field, owing to the normal conducting nature of the cavity. After calibration, the sensor was placed in the superconducting cavities to measure the penetrating DC magnetic field, by measuring the magnon frequency as the external DC magnetic field was varied. A detailed description of the setup for these kinds of measurements has been covered in past work \cite{pastWork1, pastWork2}. Loop probes were inserted into the cavity and coupled primarily to the cavity mode. The magnon mode was measured in the dispersive regime indirectly through its coupling to the $TM_{0,1,0}$ cavity mode. These measurements were thus undertaken with minimal alterations to a cavity setup used for other purposes, with only the addition of a small LiFe sphere. The external field was supplied by a commercial superconducting magnet with $\pm0.1\%$ field homogeneity over 1cm DSV from the field centre, where the cavities were placed.

Methods of detecting DC magnetic fields at cryogenic temperatures already exist. Flux gate magnetometers, for example, are a popular choice and have been used to detect DC fields external to microwave superconducting cavities\cite{FluxGateMag, ExpelFlux1, ExpelFlux2}. One could implement such methods to measure the internal fields, however, other than measuring field close to the walls, this requires the insertion of a conductor into the cavity which would prevent simultaneous measurements of the cavity mode. This would be the case for many commercial magnetometers, especially those that rely on conductive pickup loops. The main benefit of our method was that it was relatively noninvasive. Due to the high spin density and low loss of LiFe\cite{CMP_Life}, the sphere can be extremely small, with a small low-loss mount (alumina) and still couple visibly to the cavity mode in the dispersive regime. This allows accurate simultaneous measurements of the cavity mode and DC field anywhere in the cavity. LiFe also has a relatively low number of spurious modes visible at low temperatures in transmission data relative to the most common choice, Yttrium Iron Garnet (YIG)\cite{Magnonqed1}, making calibration simpler. Dielectric and magnetic losses of the materials involved in principle can degrade the cavity Q, but  could be kept to a minimum using small spheres and low-loss supports such as sapphire\cite{sapphireHighQ1,sapphireHighQ2}. A more detailed discussion, as well as some estimated losses, is present in the supplementary material.

The orientation of the crystal axis of the LiFe sphere relative to the external magnetic field was maintained between calibration and experiments, as it affects $B_\text{off}$ in equation (\ref{FtoB}). Calibration measurements of the LiFe sensor were undertaken in a Copper cavity of identical dimensions to the cylindrical Niobium cavity. The measurements revealed two similar magnon modes, which needed to be taken into account when calibrating measurements. The results of the fitting the two modes gave $g_\text{eff}=1.90$ and $B_\text{off}=0.034$\hspace{0.2em}T for the first mode and $g_\text{eff}=1.97$ and $B_\text{off}=0.019$\hspace{0.2em}T for the second. Observations of transmission of the magnon modes, were used to identify the correct calibration curve. To improve signal to noise ratio, static features were removed from transmission plots by subtracting at each frequency the average transmission over all magnetic field values prior to the first appearance of the magnon mode. The LiFe sphere used in this experiment was a $0.58$\hspace{0.2em}mm diameter sphere, mounted via some epoxy, on a $0.72$\hspace{0.2em}mm diameter, $4$\hspace{0.2em}mm tall, alumina cylindrical mount, where the (111) axis of the crystal was oriented along the direction of the mount. More detailed analysis of the spectra of the LiFe sphere used, can be found in past work \cite{CMP_Life}. It should be noted mode softening occurs near the saturation field of the sphere, thus the calibration was done between $5-22$\hspace{0.2em}GHz, limiting the sensor's operation to $B_{int}\geq0.154$\hspace{0.2em}T.

%\section{Niobium Cylindrical Cavity}\label{cylinder}

%\subsection{Cavity Geometry}
After calibration, the first superconducting cavity measured was a cylinder of internal diameter $10$\hspace{0.2em}mm and height $8$\hspace{0.2em}mm, corresponding to a fundamental $TM_{0,1,0}$ mode frequency at $22.45$\hspace{0.2em}GHz. The base and walls were cut out of a single piece of Niobium with $3$\hspace{0.2em}mm walls and a $4$\hspace{0.2em}mm base, including probe holes and a mount for the LiFe sphere. A $3$\hspace{0.2em}mm thick lid was made from a second piece of Niobium secured using six evenly spaced $6$\hspace{0.2em}mm M2 bolts. See fig. \ref{cylndraw} for a sketch. 

\begin{figure}[t!]
	\includegraphics[width=0.8\columnwidth]{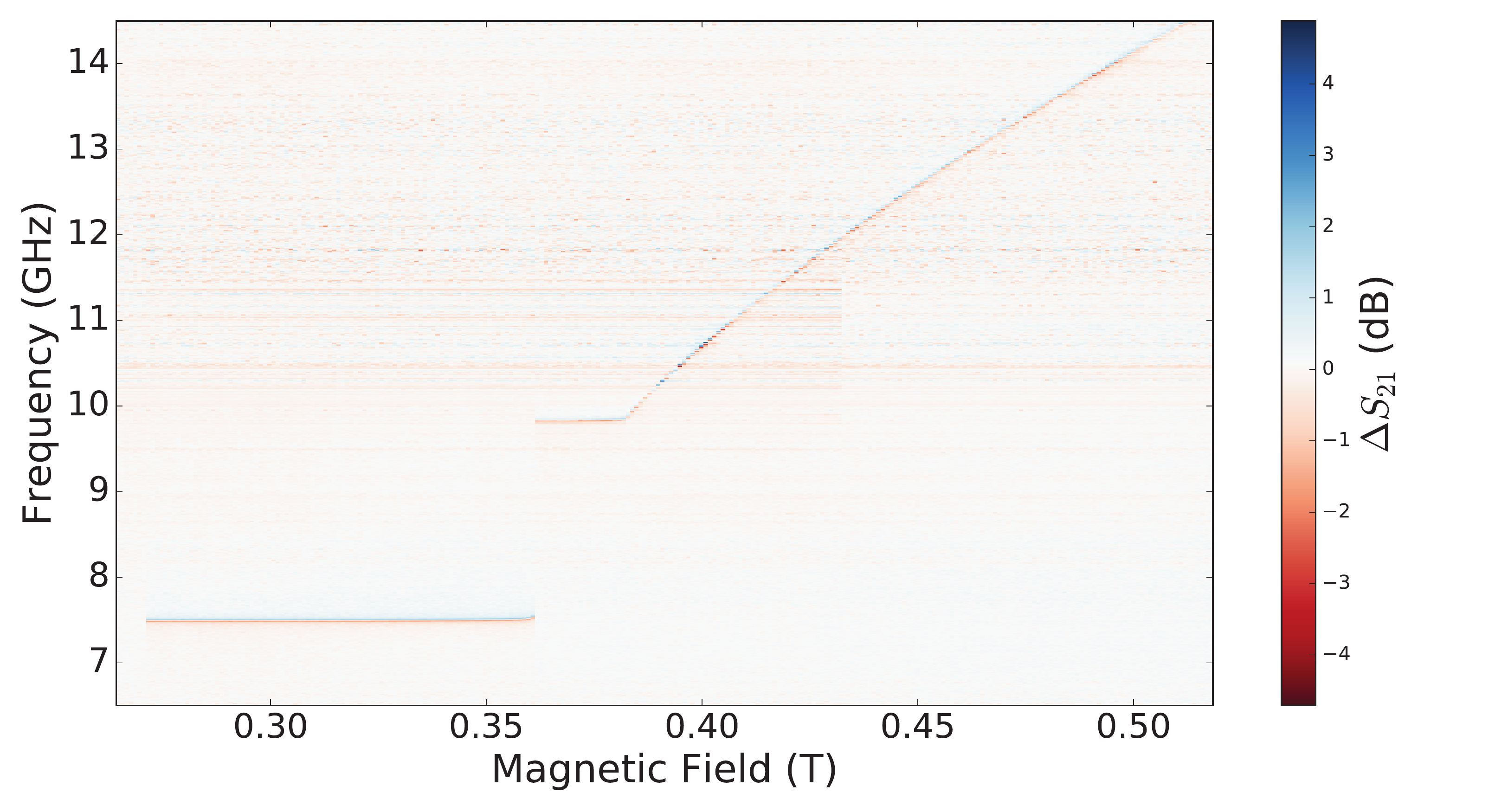}
	\caption{A filtered relative transmission of the Niobium cavity on ramp up of the magnetic field and weakly coupled probes.}
	\label{Nb_transmission}
\end{figure}

\begin{figure}[t!]
	\includegraphics[width=0.8\columnwidth]{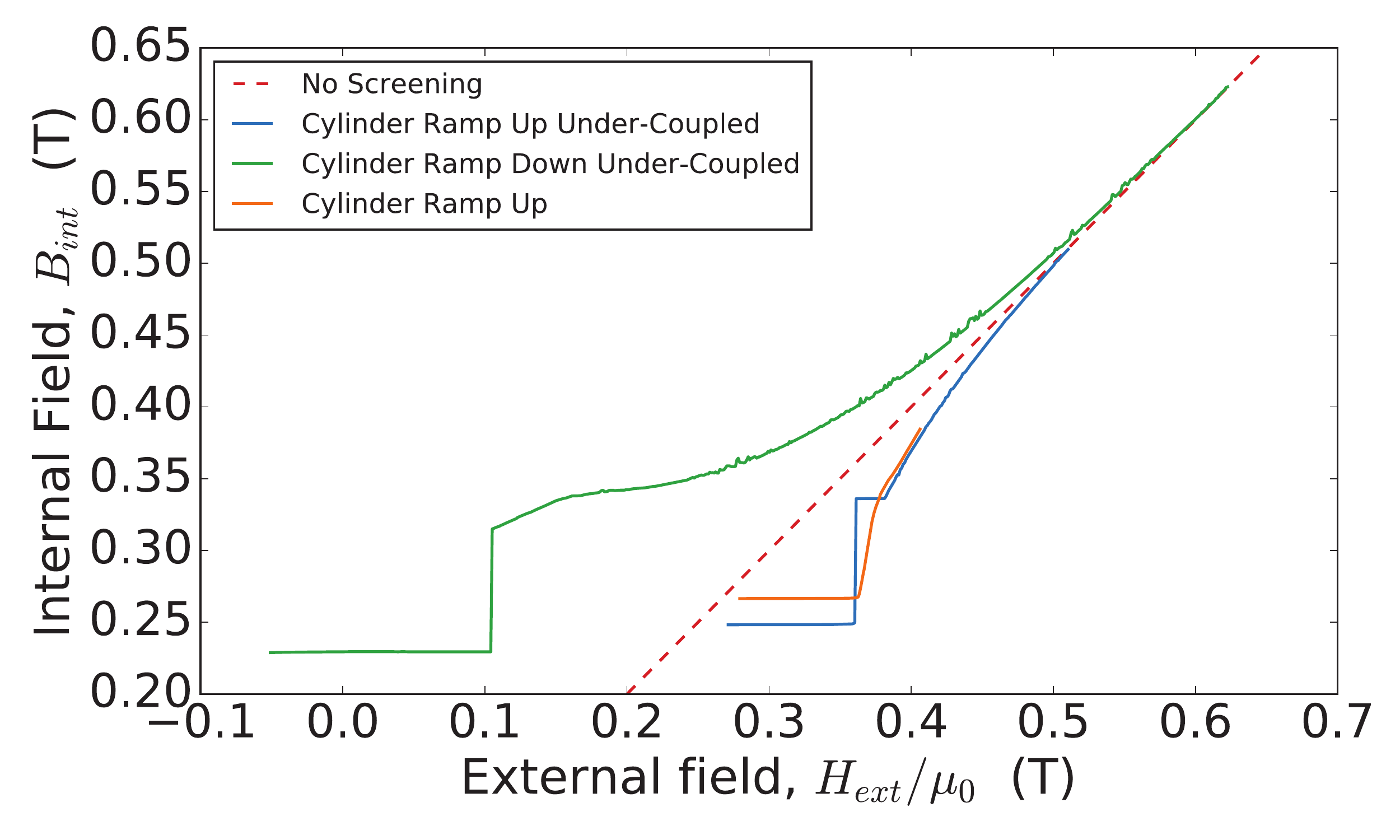}
	\caption{Cylindrical cavity internal magnetic field against external magnetic field for the ramp up and down as well as different probe configurations.}
	\label{cylnscreen}
\end{figure}

%\subsection{Measurements and Discussion}
Both the cavity and magnon modes were measured to determine how the penetrating field and superconducting material losses changed with varying DC magnetic field at $4.5$\hspace{0.2em}K in temperature. From the calibration data, the frequency of the magnon mode determined the internal magnetic field for several experimental runs. This included ramping the field up and down with the probes weakly coupled to the cavity mode (and thus even more weakly to the magnon mode), as well as ramping up the field with the probes inserted such that they are coupled directly to the magnon mode. These results are shown in fig. \ref{cylnscreen} and an example of the spectroscopy, corresponding to ramping up the magnetic field with weakly coupled probes, is shown in fig. \ref{Nb_transmission}.


\begin{figure}[t!]
	\includegraphics[width=0.8\columnwidth]{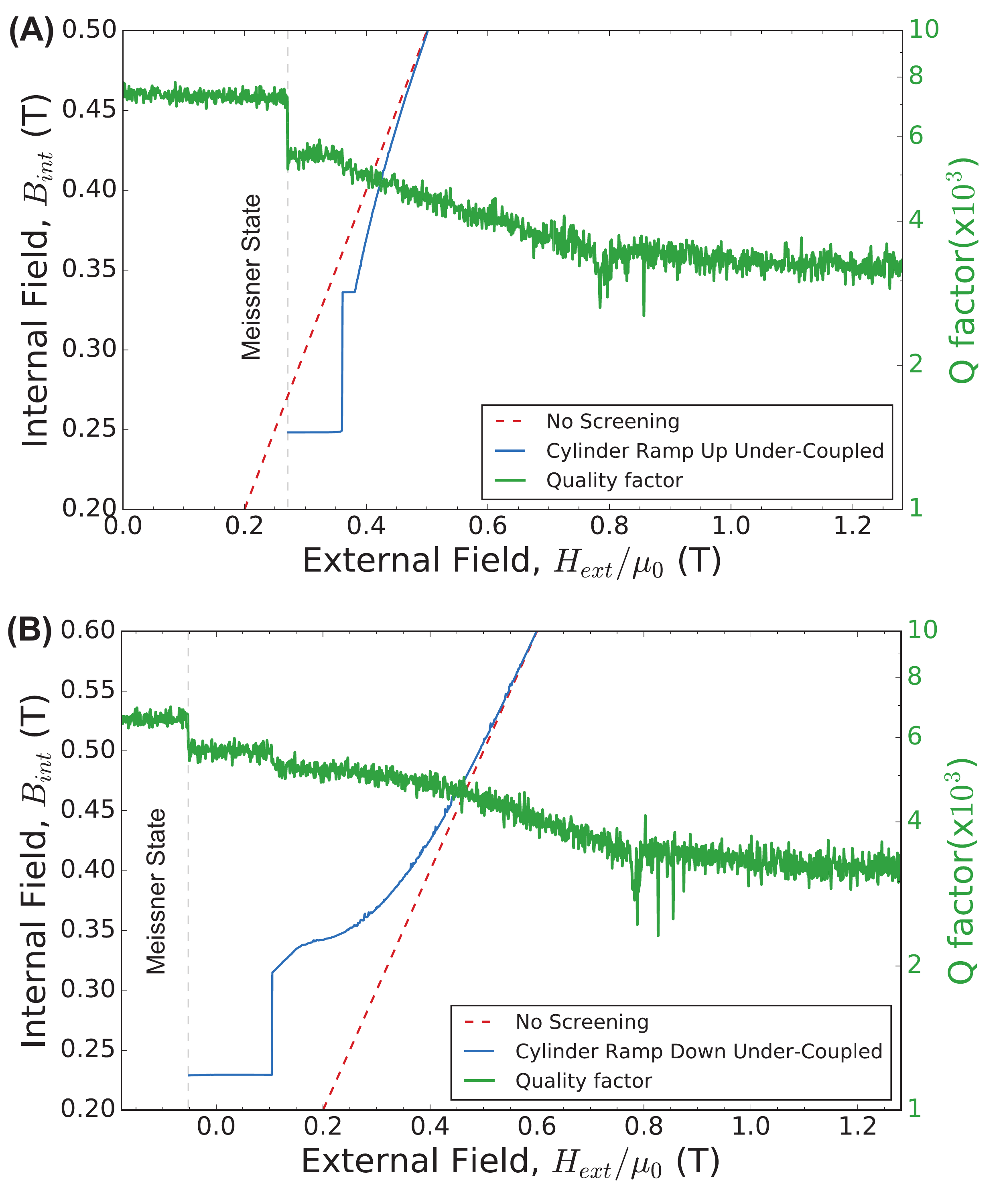}
	\caption{(A) Cylindrical cavity quality factor and screening for the under-coupled ramp up case. Measured $H_{sh}/\mu_0=0.2712\pm0.0006$\hspace{0.2em}T (B) Cylindrical cavity quality factor and screening for the under-coupled ramp down case.}
	\label{updownqs}
\end{figure}

In the case where the probes were coupled directly to the magnon mode, they are over-coupled to the cavity mode such that cavity intrinsic quality factors are no longer reliably measurable. Thus fig. \ref{updownqs}, only shows the quality factor of the cavity mode for the ramp up and down of field in the under-coupled case.

From fig. \ref{cylnscreen}, it can be seen that some of the behaviour of the measured field was as expected. On the ramp up, field enters the cavity at a consistent critical field. The field at which it was energetically favourable for vortices to form in the material was the first critical field, $H_{c1}$. However, well above this field the system can remain in a meta-stable Meissner state whereby the energy barrier associated with the forming of a vortex \cite{SurfaceBarrier,superheatingNb1} prevents the system from entering the mixed phase. This is analogous to surface tension preventing a liquid-gas transition in super-heated water. The meta-stable Meissner state, can decay due to thermal activation, tunnelling, and is affected by anisotropies and disorder \cite{superheatingNb1,VortexNucleation}. The Meissner state will become unstable at a higher field called the superheating field, $H_{sh}$.  Past theory and experiments has shown the Meissner state in pure Niobium above $H_{c1}$ to be extremely stable, such that the point at which field enters the cavity for most applications will be as large as $H_{sh}$\cite{superheatingNb1, superheatingNb2}.

The measured transition at which field enters the cavity was therefore interpreted as the superheating field. This corresponds to $H_{sh}/\mu_0=0.2712\pm0.0006$\hspace{0.2em}T for the cavity under-coupled case, which was identical to the value that can be obtained from the first drop in cavity quality factor from fig. \ref{updownqs}, and $H_{sh}/\mu_0=0.2793\pm0.0006$\hspace{0.2em}T for the case of directly coupling to the sphere. The discrepancy between the measured transition field could come from the fact that different runs have likely small differences in temperature and different magnetic field ramp rate. The measured field was initially screened by the walls and eventually returns to the un-screened case at large field. The presence of vortex motion explains why changes in field correspond to changes in cavity loss in the mixed phase. Hysteric behaviour was observed as the ramp down measurements show more field internally than provided by the external magnet. This is likely due to the vortices being trapped on the ramp down, and is consistent with fig. \ref{updownqs} (B), where the quality factors were observed to return to a lower value on the ramp down. A dip in the quality factor of the cavity mode was observed around $0.8$\hspace{0.2em}T, this was simply due to a cavity mode interaction with the magnon mode, which was tuning through at this field value.

\begin{figure}[t!]
	\includegraphics[width=0.7\columnwidth]{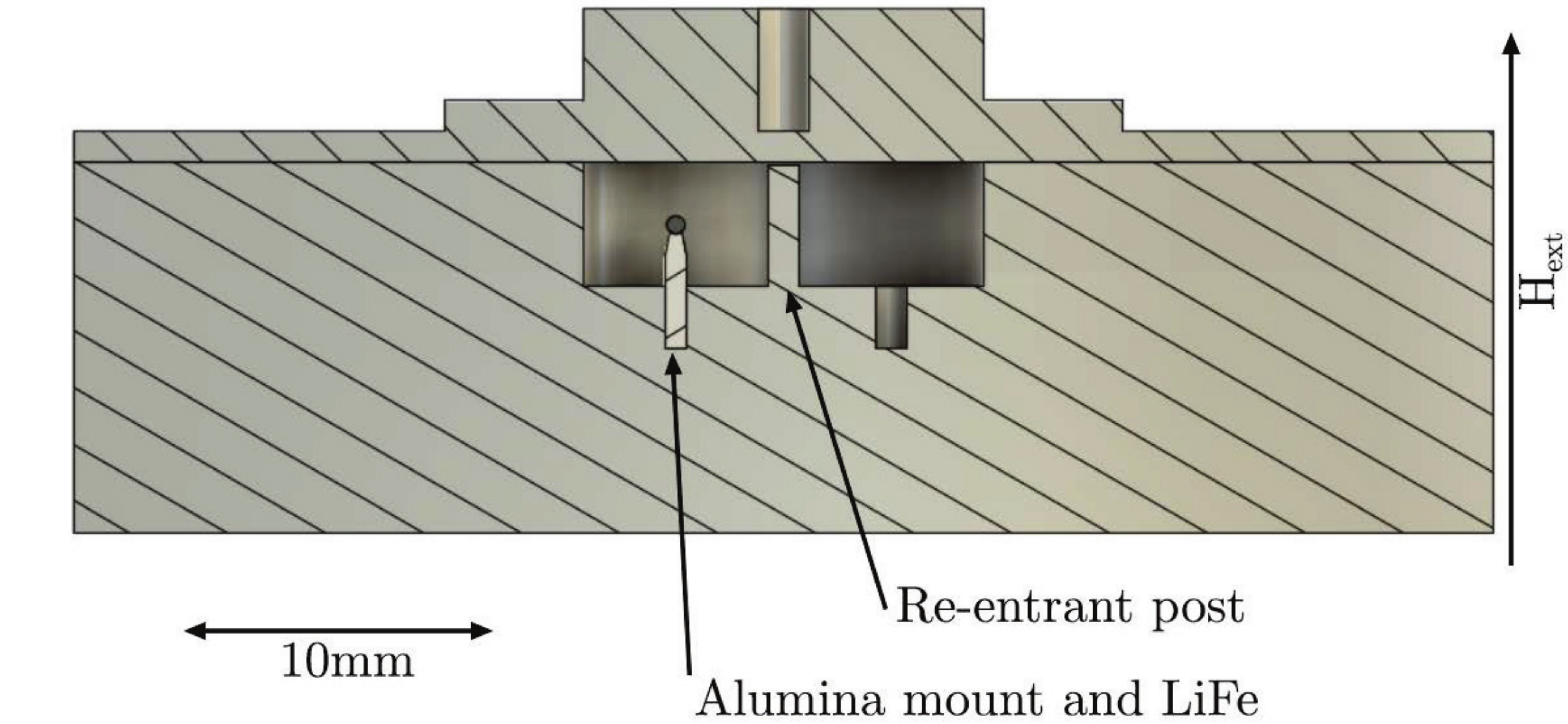}
	\caption{Cross-section of the Nb re-entrant cavity.}
	\label{reentdraw}
\end{figure}

Significant deviations from expectations were observed in the form of plateaus visible in fig. \ref{cylnscreen}. In this case, rather than a smooth transition, a constant internal field was measured as the external field was ramped, corresponding to a constant quality factor of the resonant cavity. On further ramping up, the system sees another transition as the measured field jumps at $H_{ext}/\mu_0=0.3613\pm0.0006$\hspace{0.2em}T, corresponding to the end of a plateau in loss. It is unlikely the cavity stops creating new vortices over this field range, as above $H_{sh}$ the surface barrier preventing field penetration has vanished\cite{superheatingNb1}. Thus, the plateaus are most likely due to the LiFe sphere measuring the flux associated with a fixed number of vortices in its vicinity. There are several expected transition fields between the first and second critical fields which could account for the observed behaviour. If the Nb cavity was pure enough to exhibit an IMS, a localized cluster of vortices near the sensor could explain the observed plateau. Alternatively, the transition field, at the end of the plateau in field, could correspond to the flux depinning field \cite{vortexMatter1}, in which the vortex matter transitions from solid to liquid. This explanation is consistent with the loss data as vortex pinning prevents a significant loss mechanism from vortex flow, with losses increasing after the flux depinning field with newly created vortices. The plateau in field could therefore be explained by a fixed number of vortices with magnetic field lines pinned in the vicinity of the sensor. Further work, including spacial characterization of the field, would need to be done to explore this hypothesis. Characterization of the material properties of the sample, including $\kappa$, would be necessary to determine if the sample could exhibit an IMS. A SANS experiment would also be a benefit to determine vortex matter structure \cite{vortexMatter1,SANS1}.

For this particular cavity, the probes were inserted from the bottom. This means the probe holes were aligned with the external magnetic field, and could thus play a role in behaviour seen. Intuitively, the system should favour threading flux through the empty hole rather than creating vortices, and would localize flux near the centre of the cavity. This is something that can be quantified, however. The energy associated with flux through a loop of superconductor is minimized when an integer number of flux quanta, $\Phi_0$, is threaded through. Assuming the system is in equilibrium, the energy cost, $E_{B}$, of $n$ flux quanta is equal to the energy associated with the magnetic field\cite{annett}:
\begin{equation}
E_{B} = \frac{l}{2\mu_0 A}(n\Phi_0)^2,
\end{equation}
where $\mu_0$ is the permeability of free space, $l$ is the length of the probe holes and $A$ is their cross-sectional area. Similarly, the energy associated with a single vortex, $E_{v}$, is approximately given by\cite{annett}:
\begin{equation}
E_{v} \approx \frac{l \Phi_0^2}{4\pi \mu_0 \lambda^2}\ln(\frac{\lambda}{\zeta}),
\end{equation}
where $\lambda$ is the London penetration depth, and $\zeta$ is the coherence length. Using values for these constants from the literature \cite{Poole} and the probe hole geometry, it is possible to estimate the point at which creating a vortex is energetically favourable to threading an additional flux quantum through the loop. This occurs at $n\sim10^7$, which corresponds to an average field in the probe holes of $\sim\hspace{-0.1em}0.004$\hspace{0.2em}T. Clearly, the flux through the probes holes does not play a large part in the observed behaviour.

%\section{Niobium Single Post Re-entrant Cavity}

%\subsection{Cavity Geometry}
For comparison and confirmation of these results, it was decided to measure another kind of superconducting cavity, a re-entrant cavity consisting of a cylinder of internal diameter $13$\hspace{0.2em}mm and height $4$\hspace{0.2em}mm, with a post of diameter $1$\hspace{0.2em}mm and height $3.9$\hspace{0.2em}mm. This cavity supports a fundamental re-entrant mode at $5.62$\hspace{0.2em}GHz. A cylindrical hole was machined for the alumina mount of the LiFe sensor, such that the sphere sat protruding $2$\hspace{0.2em}mm into the cavity. From a previous experiment a $1$x$6$\hspace{0.2em}mm slot had been machined on the opposite side of the post. See fig. \ref{reentdraw} for a drawing. Loop probes were inserted into the cavity along the axis perpendicular to this cross-section. In this case, the post mode was close in frequency to where the magnon mode was expected to appear when flux entered the cavity, thus only a weak probe coupling configuration was necessary to resolve both the cavity and  magnon modes clearly. Due to the cavity mode frequency being located near the magnon mode's initial frequency at $H_{sh}$, results presented were restricted to the dispersive regime, to ensure the validity of the calibration, with a maximum error due to hybridization of less than $1\%$ near the cavity mode (see supplementary materials). 

%\subsection{Measurements and Discussion}
\begin{figure}[t!]
	\includegraphics[width=0.8\columnwidth]{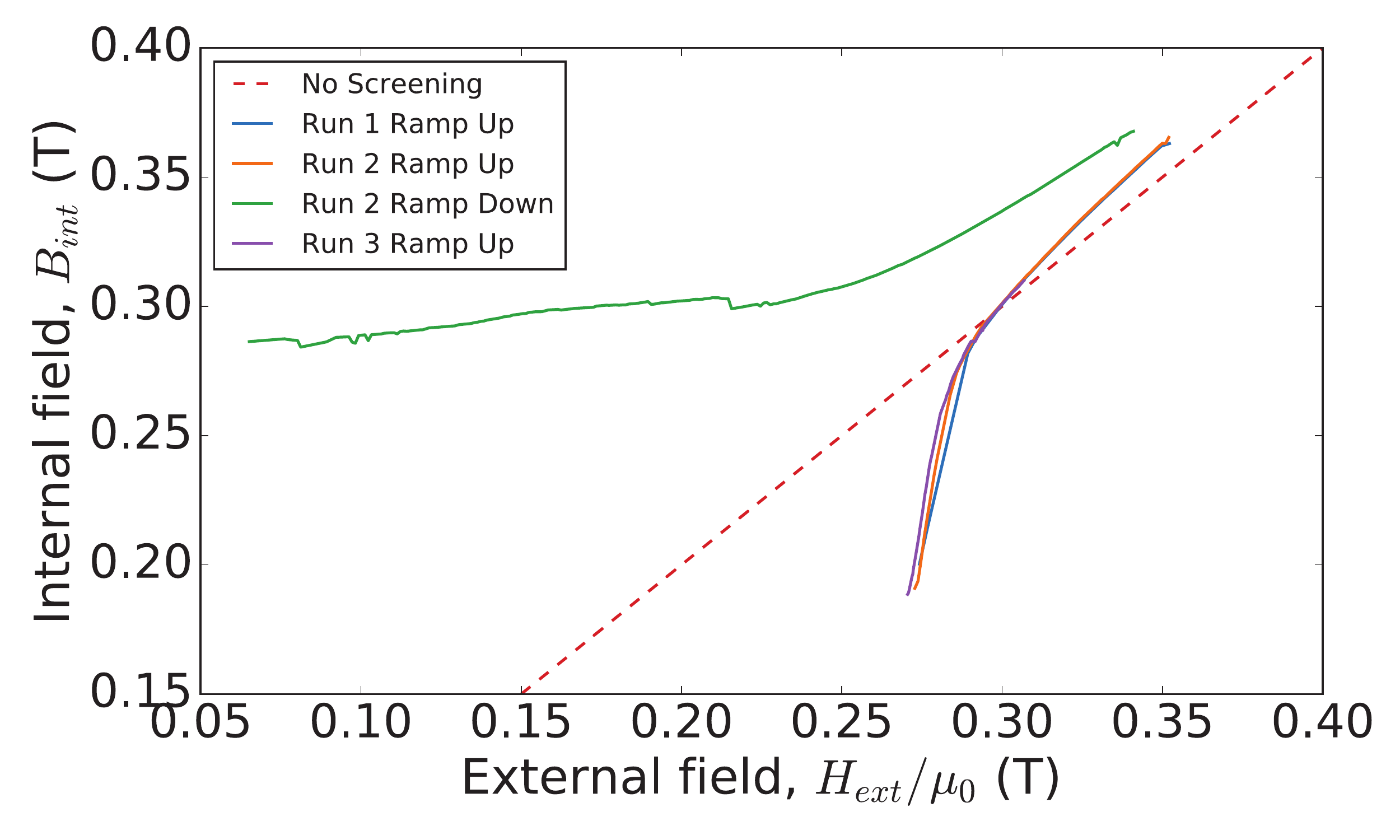}
	\caption{Re-entrant cavity internal magnetic field against external magnetic field for both the ramp up and down. Measured $H_{sh}/\mu_0=0.272\pm0.001$\hspace{0.2em}T}
	\label{reentscreen}
\end{figure}

The results of measured internal against external field are shown in fig. \ref{reentscreen}, where the first appearance of the magnon mode provides a consistent measurement of $H_{sh}/\mu_0=0.272\pm0.001$\hspace{0.2em}T. These results show that the field measured inside the cavity becomes higher than the field outside, even on the ramp up. This can potentially be explained by the presence of the post inside the cavity acting as a diamagnet and expelling field up to the second critical field. With the sphere nearby to the post, this would enhance the local field measured by the sphere, and would also explain why a plateau isn't observed, as screening from the post diverts flux from vortices over a wider area toward the sphere.  Thus if the experiment were repeated, ramping up to larger magnetic fields, we would expect the increase in internal field to eventually return to the no screening case. The usual hysteric behaviour due to trapped vortices was observed on the ramp down.\\

%\section{Conclusions}
In summary, a method of intracavity DC magnetic field sensing was tested on two Niobium cavities. The screening effects of these cavities were measured, demonstrating the transition from perfect diamagnetism by the Meissner effect, to negligible screening effects when measured above the second critical field. Both cavities measured a consistent transition field for DC magnetic field to enter the material, interpreted as the superheating field, and showed hysteric behaviour for increasing and decreasing field due to trapped vortices. Some unexpected behaviour occurred in these cavities. The cylindrical cavity demonstrated some observed plateaus just above the first critical field, the origin of which are unknown, however could be due to a transition in the structure of vortices in the material. The re-entrant cavity did not demonstrate the same plateaus, however did see an enhancement of field on the LiFe sphere over some range, most likely due to the post screening field. Possible future investigations include observing the decay of persistent fields by this method, measurement of the penetrating field as it varies spatially within the cavity, and the effect of cavity geometry including the use of thin films compared to the bulk case here.

\section*{Acknowledgements}
This work was supported by the Australian Research Council grant number DP190100071 and CE170100009 as well as the Australian Government's Research Training Program.
\section*{Data availability statement}
The data that support the findings of this study are available from the corresponding author
upon reasonable request.

%\section*{References}
\bibliography{refs}
\renewcommand{\theequation}{S\arabic{equation}}
\renewcommand{\thetable}{S\Roman{table}}

\begin{figure*}[t!]
	\begin{minipage}{2.1\columnwidth}\flushleft
		\textsf{\textbf{\Large Determination of Niobium Cavity Magnetic Field Screening via a Dispersively Hybridized Magnonic Sensor - Supplementary Material}}
	\end{minipage}
\end{figure*}

\newpage
\section*{}\newpage
To help recreate the sensor setup, provided in table \ref{tabParams} are cavity-magnon coupling parameters for the two cavities, as well as magnon line-widths. Cavity probe coupling parameters were not measured directly, however, similar probe couplings can be replicated based on the cavity quality factors and transmission ($|S_{21}|$) on resonance (also provided in the table).

\begin{table}[h]
	\begin{tabular}{| l | l | l |}\hline
		System Parameters     		       & Cylindrical cavity & Re-entrant   \\\hline
		Cavity-magnon coupling, $g_{cm}$   & $55$\hspace{0.2em}MHz            & $120$\hspace{0.2em}MHz*    \\\hline
		Magnon line-width    		       & $7.5$\hspace{0.2em}MHz           & $7.5$\hspace{0.2em}MHz     \\\hline
		Meissner state cavity Q factor     & $7780$      	   & $110$    \\\hline
		Meissner state cavity $|S_{21}|$  & $-57$\hspace{0.2em}dB      	    & $-62$\hspace{0.2em}dB     \\\hline
		\multicolumn{3}{l}{\footnotesize *Theoretical estimate}
	\end{tabular}
	\caption{System parameters for the cylindrical and re-entrant cavities.}
	\label{tabParams}
\end{table}

Operating the sensor such that the magnon mode is detuned from the cavity mode on the scale of the cavity-magnon coupling, $g_{cm}$ will result in hybridization, shifting the position of the measured mode. This can be easily quantified using coupled mode theory \cite{cavitymagnon}:
\begin{equation}
\omega_{\pm} = \frac{\omega_c+\omega_m}{2} \pm \sqrt{\frac{(\omega_c-\omega_m)^2}{4}+g_{cm}^2},
\end{equation}
where $\omega_{\pm}$ corresponds to the hybridized mode frequencies, and $\omega_m$ and $\omega_c$ are the magnon and cavity mode frequencies respectively. The absolute frequency error associated with hybridization can be determined as the difference between hybrid mode in the calibration cavity compared to that of the measured cavities. For the cylindrical cavity, there is no error due to hybridization as the calibration was done in an identical cavity. In the re-entrant cavity, it corresponds to $|\omega_{re,+}-\omega_{cal,-}|$, where $\omega_{re,+}$ is the positive branch of the measured re-entrant cavity, $\omega_{cal,-}$ is the negative branch of the calibration and the same $\omega_m$, calculated based on the equation (1) in the main text, is used for both hybrid models.\\

For the re-entrant cavity the magnon mode first appeared near the cavity mode at $H_{sh}$. It was, therefore, difficult to measure the cavity-magnon coupling strength and a theoretical estimate was used instead. Measured and theoretical values are then used to determine regions of operation that would produce larger than $1\%$ error from hybridization. This corresponded to $\pm250$\hspace{0.2em}MHz around the $5.62$\hspace{0.2em}GHz cavity mode. Data taken within this range was dropped from the results in fig. 6 in the main text. Superheating field can still be reliably quoted from this region as external measured field is unaffected. No other cavity modes were near enough to the magnon mode to be relevant.\\

For high performance cavities, including lossy dielectrics or magnetic materials would no doubt degrade performance. These losses are related to electric and magnetic filling factors of the materials respectively, and therefore will contribute differently based on cavity geometry. The alumina mount is not expected to degrade our performance at 4K, although will be unsuitable for high Q cavities at milliKelvin temperatures due to increased loss\cite{aluminamK}. For lower loss, a sapphire\cite{sapphireKrupka} or even PTFE\cite{PTFEKrupka} mount could be used. Unfortunately little data is available for the microwave losses at cryogenic temperatures for LiFe. Yttrium Iron Garnet (YIG), the most common choice of material in magnon experiments, is expected to be lower loss, and an upper estimate on losses can be used to demonstrate a similar suitability for high Q cavities\cite{Magnonqed1}. These losses can be quantified by filling factors and loss tangents of the material\cite{PTFEKrupka}:
\begin{equation}
Q_i = \frac{1}{p_i\tan\delta_i},
\end{equation}
where $Q_i$ is the quality factor associated with the material, $p_i$ is the electric or magnetic filling factor\cite{PTFEKrupka,Magnonqed1} and $\tan\delta_i$ is the electric or magnetic loss tangent of the material. Q factors from different loss mechanisms are added as $Q_{total}^{-1} = \sum_iQ_i^{-1}$. Alumina has a dielectric loss tangent of $\tan\delta_e\approx1\times10^{-5}$ at $4$\hspace{0.2em}K and $20$\hspace{0.2em}GHz\cite{aluminamK}. Sapphire can have a dielectric loss tangent of $\tan\delta_e\lesssim2\times10^{-8}$ at $13$\hspace{0.2em}GHz and mK temperatures\cite{sapphireHighQ1,sapphireHighQ2}. PTFE has loss tangent, $\tan\delta_e=2\times10^{-6}$ at $18$\hspace{0.2em}GHz and  $25$\hspace{0.2em}K\cite{PTFEKrupka}. Without data on the losses of LiFe at cryogenic temperatures and microwave frequencies it is difficult to produce similar estimates on Quality factor. However, we can use the Quality factors measured in past work\cite{Magnonqed1} to place a rough upper bound on dielectric losses of YIG, which would be a simple substitute in this experiment. The upper estimate of YIG losses gives $\tan\delta_e\leq9\times10^{-4}$, as filling factors were of order $0.1-1$. Whilst YIG is expected to have lower microwave loss than LiFe, with some chemical substitutions, LiFe's loss can be greatly reduced even at room temperature\cite{LiFeLowLoss}. Finally, the use of adhesive to affix the sphere to the mount, which can also be a source of loss, can be forgone by using a larger mount.\\

\begin{table}[h!]
	\begin{tabular}{| l | l | l | l |}
		\hline
		& \multicolumn{2}{l|}{Cylindrical cavity} & \multirow{2}{*}{Re-entrant} \\ \cline{1-3}
		\textbf{Mount} & TM010                        & TM011                    &                          \\ \hline\hline
		Alumina $p_e$     & $3.4\times10^{-1}$        & $9.8\times10^{-2}$       & $2.9\times10^{-4}$        \\ \hline
		Sapphire $p_e$    & $3.7\times10^{-1}$        & $1.1\times10^{-1}$       & $3.2\times10^{-4}$        \\ \hline
		PTFE $p_e$& $9.5\times10^{-2}$        & $1.8\times10^{-3}$       & $1.1\times10^{-4}$        \\ \hline
		Alumina $Q_e$     & $2.9\times10^{5}$  		  & $1.0\times10^{6}$        & $3.4\times10^{8}$         \\ \hline
		Sapphire $Q_e$    & $\gtrsim1.4\times10^{8}$  & $\gtrsim4.5\times10^{8}$ &  $\gtrsim1.6\times10^{8}$ \\ \hline
		PTFE $Q_e$        & $5.3\times10^{6}$ 		  & $2.8\times10^{8}$        & $4.5\times10^{9}$         \\ \hline
	\end{tabular}
	\caption{Calculated electric filling factors and Q factors associated with material losses of the sensor mount.}
	\label{tabLossMount}
\end{table}

\begin{table}[h!]
	\begin{tabular}{| l | l | l | l |}
		\hline
		& \multicolumn{2}{l|}{Cylindrical cavity} & \multirow{2}{*}{Re-entrant} \\ \cline{1-3}
		\textbf{Sphere}   & TM010                     & TM011                    &                            \\ \hline\hline
		YIG $p_e$         & $2.0\times10^{-2}$        & $5.6\times10^{-4}$       & $2.5\times10^{-5}$         \\ \hline
		YIG $Q_e$         & $\gtrsim5.6\times10^{4}$  & $\gtrsim2.0\times10^{6}$ & $\gtrsim4.4\times10^{7}$   \\ \hline
	\end{tabular}
	\caption{Calculated electric filling factors and Q factors associated with material losses of the ferrimagnetic sphere (assuming an alumina mount).}
	\label{tabLossSphere}
\end{table}

It should be noted that filling factor scales inversely with the size of the cavity, and the cavities tested here were much smaller than those typically applied in axion haloscopes or linear accelerator experiments. Similarly, electric filling factor could be further reduced by placing the sensor at a node in the electric field of the modes of interest. In this experiment, this is the case for the $TM_{0,1,1}$ mode of the cylindrical cavity, which still couples strongly to the magnon mode. Filling factors for the cavities used were calculated based on simulations using COMSOL Multiphysics\textsuperscript{\copyright} of the measured $TM_{0,1,0}$ (high electric filling factor) and unmeasured $TM_{0,1,1}$ (low electric filling factor) modes of the cylindrical cavity, as well as the measured post mode of the re-entrant cavity. Results of dielectric losses for our cavities, based on these simulations, are presented in Table \ref{tabLossMount} for the mount and Table \ref{tabLossSphere} for the ferrimagnetic sphere. This comparison provides a guide for adapting the sensor for a high performance cavity.

\end{document}